      [1999/12/01 v1.4c Il Nuovo Cimento]
\documentclass{cimento}
\usepackage{graphicx}

\title{UHECR bending, clustering and decaying feeding \\
  gamma anisotropy}
\author{D.~Fargion\from{ins:x}, \from{ins:evil}D.~D'Armiento\from{ins:x},P.~Paggi\from{ins:x}
}
\instlist{\inst{ins:x} Physics Department, Rome University 1, Sapienza, Ple.A.Moro 2,Rome
  \inst{ins:evil} INFN, Rome 1, Italy}

\begin{document}

\maketitle

\begin{abstract}
Ultra High Cosmic Rays)  made by He-like lightest nuclei might  fit clustering along Cen A. Moreover He like UHECR nuclei explain Virgo  absence because the light nuclei fragility and opacity above a few Mpc. We foresaw (2009) that UHECR He from Cen-A AGN being fragile should partially fragment into secondaries  at  tens EeV  multiplet (D,$He^{3}$,p) as the recent twin multiplet discovered ones (AUGER-ICRC-2011), at $20$ EeV  along Cen A UHECR clustering.  We suggest that UHECR are also heavy radioactive galactic nuclei  as $Ni^{56}$,  $Ni^{57}$  and $Co^{57}$,$Co^{60}$ widely bent (tens degree up to $\geq 100^{o}$)  by galactic fields.  UHECR radioactivity (in $\beta$ and $\gamma$ channels) and decay in flight at hundreds keV  is boosted (by huge Lorentz factor $\Gamma_{Ni}\simeq 10^{9}- 10^{8}$)  leading to PeVs electrons and consequent synchrotron TeVs gamma offering UHECR-TeV correlated sky anisotropy. Electron and tau  neutrinos secondaries at PeVs maybe the first signature of such expected radioactive secondary tail.
\end{abstract}

\section{Introduction}
\begin{figure}
\includegraphics[angle=0,scale=.14]{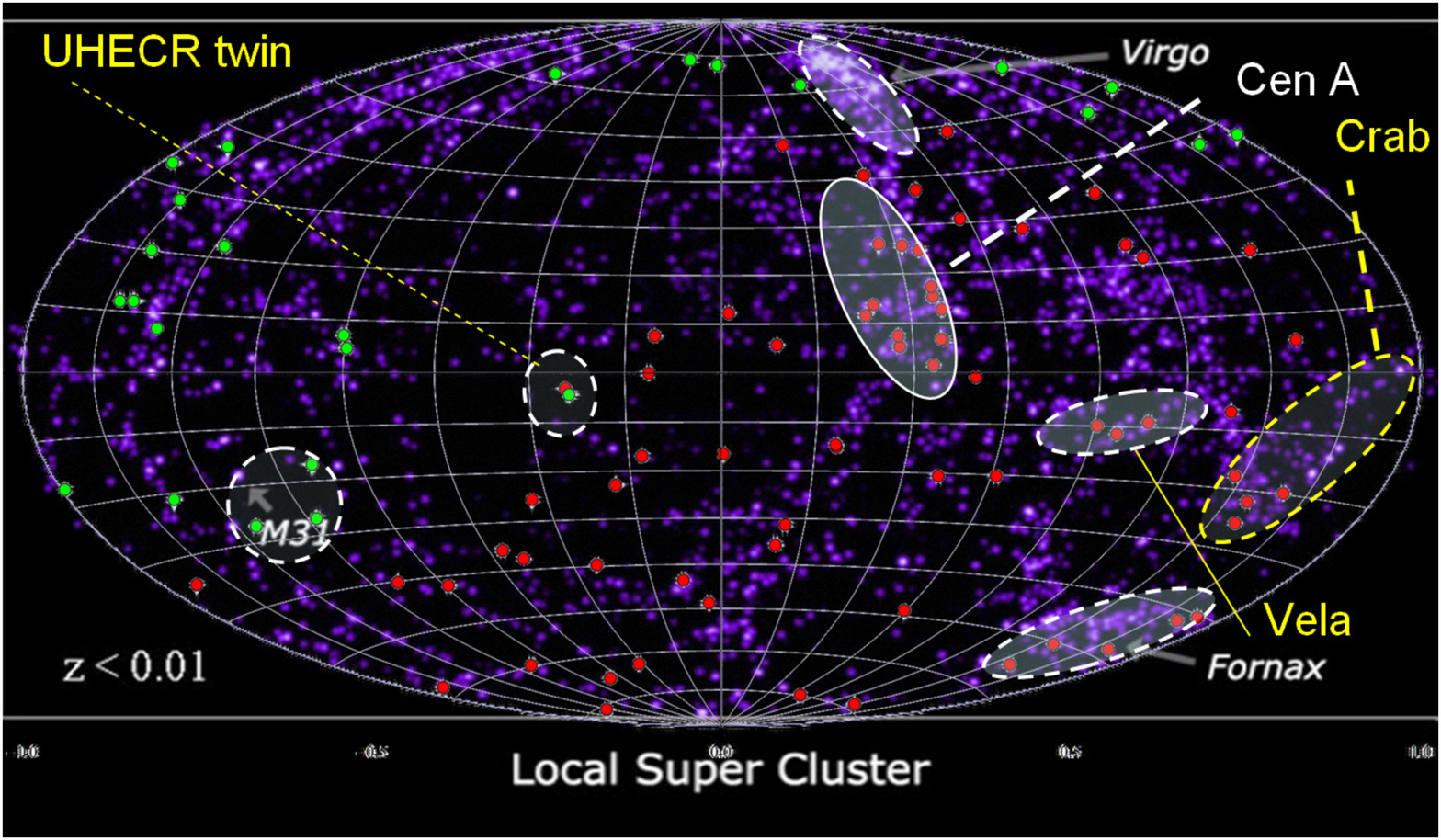}
\includegraphics[angle=0,scale=.15]{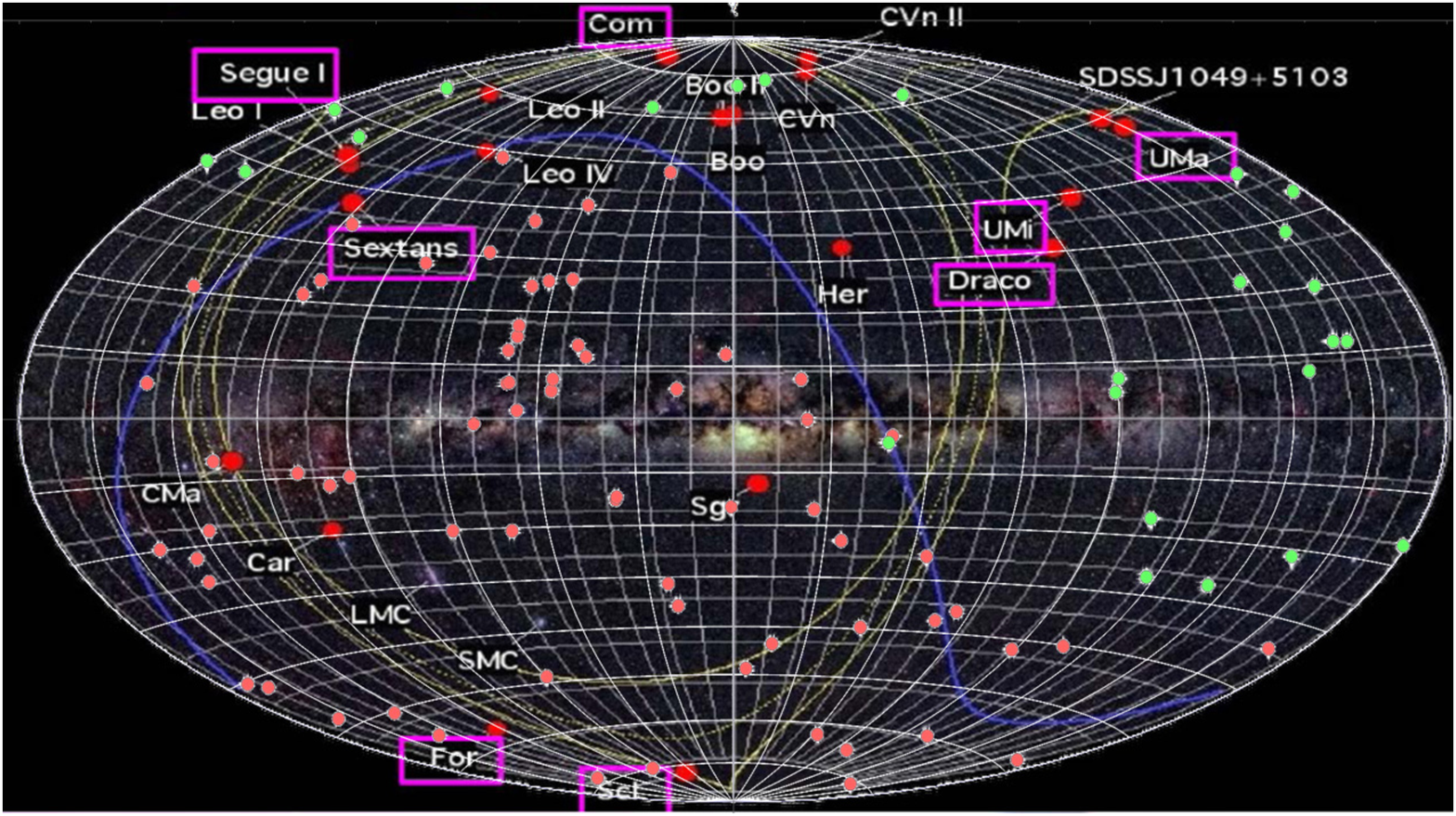}
\caption{Left: Recent  extragalactic nearby (below few tens Mpc) micron map of nearby infrared sources, versus the 69 UHECR by AUGER (red disk) and by Telescope Array (green disk) \cite{TA-12} events recently published; note the absence of Virgo, the twin UHECR AUGER-TA event on galactic plane \cite{Troitsky}, the triplet around Andromeda M31,  the clustering along Cen A and the partial clustering along the unique  nearby Dwarf Galaxy FORNAX and Sculptur. Right: the same mirror UHECR map over most of nearest Dwarf Galaxy map: note the possible role of seeds of UHECR by such nearest sources as Fornax, Sculptor as well as the eventual galactic role of Vela, Orion Clouds and others. The absence of the Galactic center is understood if heavy nuclei are widely bent by strongest magnetic fields. }\label{figure12}
\end{figure}

 Cosmic Rays (CR) origin are still a puzzle mostly because of the smearing of their arrival directions by random galactic magnetic fields.
 Just a century ago  Hess noted  and discovered the cosmic rays, CR, enigma: flying on balloon the  radioactivity first decayed but soon  grows at high altitude, probing their extraterrestrial nature. Because CR are charged their trajectories are bent and their sources are  possibly galactic and  cosmic too. At highest energy their bending, if they are nucleons, is negligible. Half a century ago, John Linsley and Livio Scarsi  have shown the existence of such  highest CR, whose tracks are almost un-deflected, offering in principle a new particle astronomy.This is the goal of UHECR astronomy \cite{Auger-Nov07}. But such  UHECR suffer also of cosmic opacity due to  cosmic MWB radiation discovered also half century ago; the opacity (a so called GZK cut-off \cite{Greisen:1966jv}) make UHECR to be nearby. Just at $1\%$ or less of cosmic size:  few tens Mpc. Along last two decade  we  were surprised  by such a first extreme Ultra High Energy Cosmic Rays, UHECR, by Fly's Eye.   However no obvious source candidate source was located in that unique Fly's Eye event direction. Such puzzling events were seen more and more for a decade again in AGASA records on 1990-2000, apparently without GZK cut off, triggering exotic models where the currier are UHECR neutrinos and the target are relic cosmic neutrinos in dark galactic halos: such Z-boson birth and decay were possibly the UHECR observed \cite{Fargion1997}. Such overabundant GZK events nevertheless almost fade away in late decade 2000-2012 by more detailed HIRES and AUGER data: these two experiments did confirm an apparent GZK cut off ; later on AUGER claimed a probable Super-Galactic correlation.  Therefore most general model claimed that those few UHECR were as expected within a GZK cut off, born by AGN and  surviving only from  nearest Universe, (one percent size) of the cosmic radius.  UHECR, if proton had to be keeping directionality because they are almost un-deflected above tens EeV energy. Just $5$ years ago Auger apparently confirmed such a GZK map observing traces of a Super Galactic Plane map by 27 events. However more  recently (2010)  69 events of UHECR data and more composition signals  disclaimed such a clear discover: indeed composition favored nuclei (whose bending is large) over  nucleons and last UHECR events diluted any apparent Super-Galactic imprint  \cite{Auger10}. HIRES and TA , on the contrary, may still favor nucleons. Therefore the disagreements are confusing any sharp understanding. Only a main  UHECR clustering along Cen A  survived AUGER recent spread map (nearly a fifth of all the events). Also  a remarkable  double  UHECR twenty-EeV multiplet is pointing  toward the same active AGN, Cen A.  We concluded since earliest $2008$ \cite{Fargion2008} and present time that lightest nuclei as He  may explain the extragalactic  AGN Cen A clustered component while explaining Virgo paucity; see fig. \ref{figure12}. UHECR He are fragile they cannot reach us from Virgo (20 Mpc) but they may  arrive from nearer (3 Mpc) Cen A.  We have foreseen ( as it is possibly it is being observing ) such UHECR He breaking into secondaries fragments at half or fourth energies; these fragments may be those observed as a train of events at twenty EeV in twin multiplet along Cen A \cite{Fargion2009-2010}, \cite{Auger11}. However majority of remaining  UHECR events might be  correlated with other Gamma and TeV anisotropy; these map correlations are suggesting also a galactic heavy nuclei UHECR component (Ni,Co), whose eventual radioactive decay in flight may explain the ARGO-ICECUBE TeV  apparent correlating map  with UHECR one.  Such tens EeV nuclei may feed, by their beta decay in flight, other secondaries as PeV electrons, later on source of synchrotron TeV gamma spread signals and neutrinos. Also UHE neutrons, fragments of He UHECR from Cen A, may shine by PeVs electrons into TeV synchrotron photons.Also inner galactic center maybe source of UHECR whose bent traces are lost, but whose neutral nucleons may still shine at EeV  energy. Fragments as PeV neutrinos discovered on May 2012 by ICECUBE may be the earliest candidate  to be such  $\nu_{e}$, $\nu_{\tau}$ showering into ICECUBE, neutrinos born by UHECR  beta decay more than expected EeV GZK ones.

\begin{figure}
\includegraphics[angle=0,scale=.12]{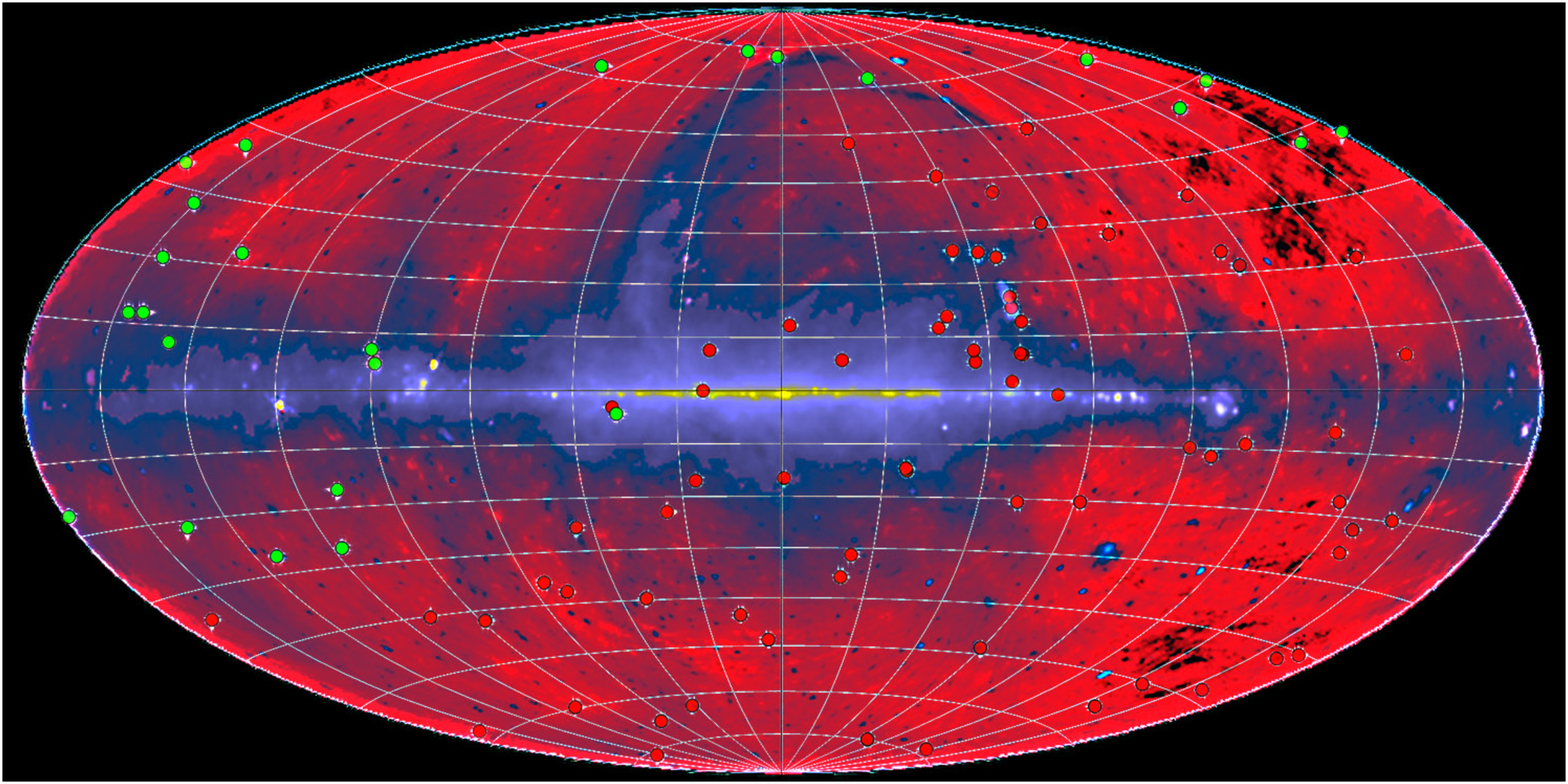}
\includegraphics[angle=0,scale=.19]{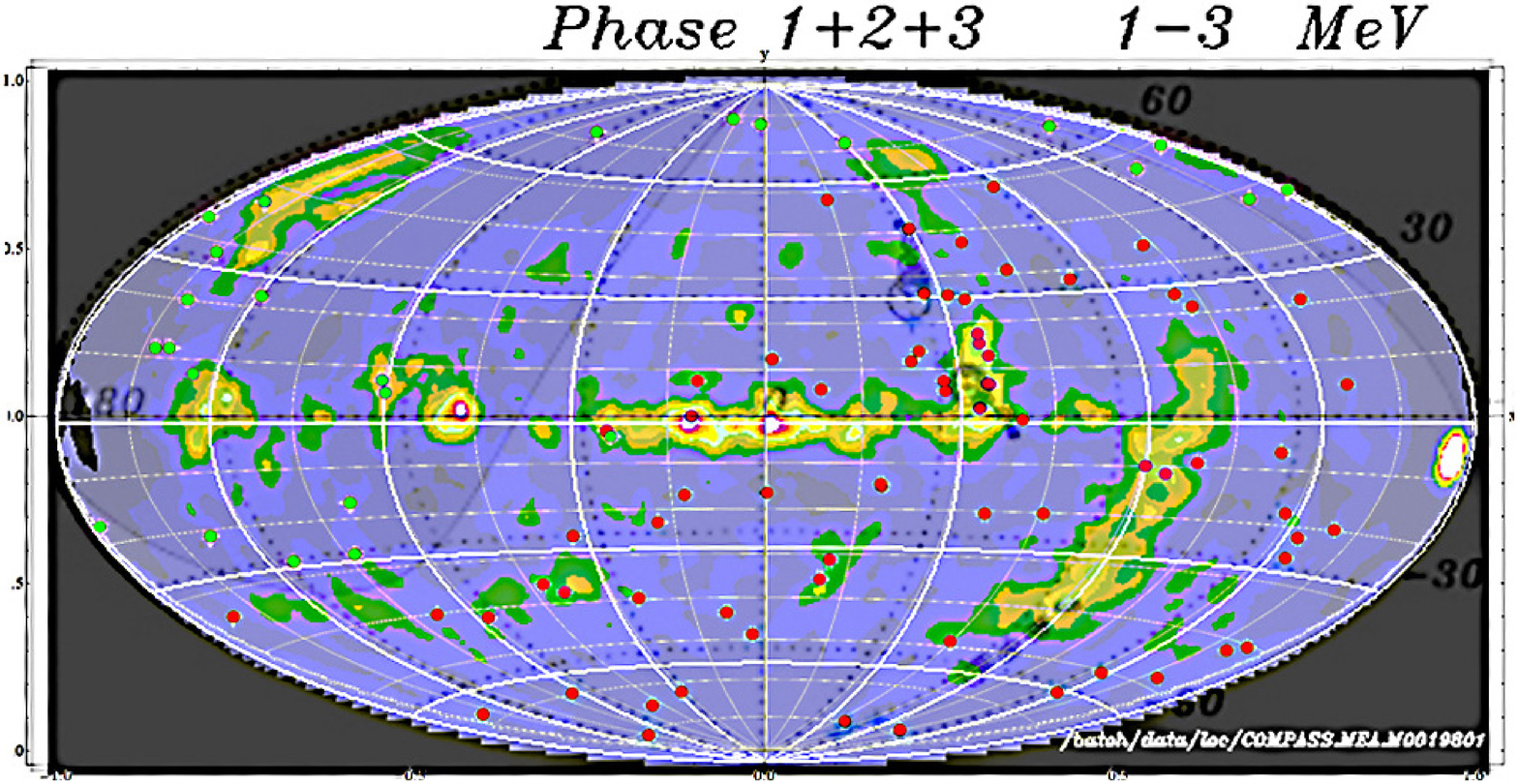}
\caption{The UHECR events over the 408 MHz radio map: note a common clustering along Cen A and Vela. The radio anisotropy may be just a trace of the sources both galactic and extragalactic ; the MeV gamma map whose tiny anisotropy follows the UHECR events may be indebt also to local galactic sources hinting for a role of Vela, Magellanic Clouds and streaming, Orion and Crab nebula. }\label{figure34}
\end{figure}

\begin{figure}
\includegraphics[angle=0,scale=.32]{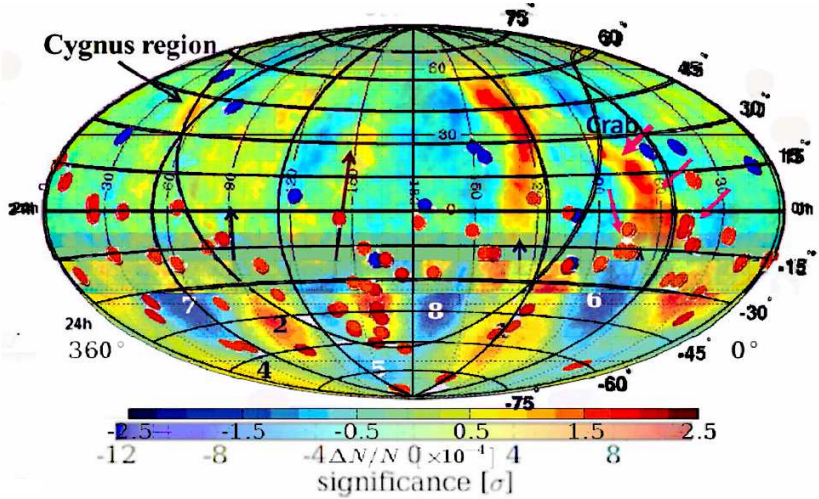}
\caption{TeV gamma Anisotropy at TeV (see \cite{ARGO},)over the UHECR by AUGER and oldest HIRES one. The Fornax region in the low right side is correlated to UHECR clustering, as well as the Vela region and an area nearby the galactic center. The Cen A clustering is also correlated to the TeV anisotropy. We foresee a correlation also with the TA events. }\label{figure5}
\end{figure}

 The common wisdom on CR teach that they are galactic up to PeVs energies (the knee) and later on they start to be dominated by extragalactic heavy nuclei contribution above the ankle (near EeV energy); at tens EeV energy most of the author believe  the UHECR are extragalactic and possibly nucleons. It is possible to overlap different maps   to trace apparent or real UHECR correlation. We believe that such a test is a key tool in disentangle the real UHECR origin and nature. One may combine AUGER with recent published TA UHECR events. It is important to note that often such maps are not in the same coordinate frame. We did updated the AUGER with the TA maps over  infrared map in galactic coordinate chosen in the GZK near Universe (see Figure below) \ref{figure12}: the absence of Virgo is suggesting a veto (as for He fragile nuclei) avoiding their arrival. The second map in a mirror galactic coordinate show the eventual correlation with nearest Dwarf galaxy and UHECR.  The Fornax and Sculptur Dwarf galaxy correlation maybe indebt to a cluster of events. As one may see in update maps the Radio 408 MHz correlation is favoring the Vela and possibly the Orion-Crab connection. The old celebrated Comptel map with both AUGER and TA events is also showing  a possible local source of the events, see fig. \ref{figure34}. Finally the presence of remarkable clustering of UHECR along Cen A, Vela and Orion -Crab region as well as along Fornax (see Fig.\ref{figure5})  is suggesting the role of UHECR radioactive heavy nuclei whose bending may trace local sources and whose composition may fit observed air-shower morphology. In this frame PeV neutrinos and tens TeV gamma anisotropy may be secondary fragments of growing data either we shall see the rise of SuperGalactic imprint or we are just discovering few local galactic sources and Cen A \cite{Fargion2011}, In conclusion we are witness of the century long puzzle solution, with the eventual birth of a correlated PeV neutrino astronomy. Possibly of tau nature at tens PeV or EeV energy band\cite{FarTau},\cite{Auger08}, \cite{Feng02},\cite{Learned}, rising as amazing up-going tau air-showers in exit from  Ande or our Earth.


\end{document}